\documentclass[amsmath,twocolumn,amssymb,prb,aps,superscriptaddress,letterpaper,floatfix,showpacs,citeautoscript]{revtex4}

\usepackage{graphicx}
\usepackage{dcolumn}
\usepackage{bm}
\usepackage{amsmath}

\newcommand{\E}{\mathrm{e}}
\newcommand{\I}{\mathrm{i}}
\newcommand{\eps}{\varepsilon}
\newcommand{\ket}[1]{| #1 \rangle}
\newcommand{\bra}[1]{\langle #1 |}

\begin{document}

\title{The Aharonov-Bohm effect in graphene rings}

\author{J{\"o}rg Schelter}
\affiliation{Institute for Theoretical Physics and Astrophysics,
University of W{\"u}rzburg, 97074 W{\"u}rzburg, Germany}

\author{Patrik Recher}
\affiliation{Institute for Mathematical Physics, TU Braunschweig, 38106 Braunschweig, Germany}

\author{Bj{\"o}rn Trauzettel}
\affiliation{Institute for Theoretical Physics and Astrophysics,
University of W{\"u}rzburg, 97074 W{\"u}rzburg, Germany}

\date{\today}

\begin{abstract}
This is a review of electronic quantum interference in mesoscopic ring structures based on graphene, with a focus on the interplay between the Aharonov-Bohm effect and the peculiar electronic and transport properties of this material. We first present an overview on recent developments of this topic, both from the experimental as well as the theoretical side. We then review our recent work on signatures of two prominent graphene-specific features in the Aharonov-Bohm conductance oscillations, namely Klein tunneling and specular Andreev reflection. We close with an assessment of experimental and theoretical development in the field and highlight open questions as well as potential directions of the developments in future work.
\end{abstract}

\pacs{
	72.80.Vp, 
73.40.Lq, 
74.45.+c, 
85.35.Ds 
}

\maketitle

\section{Introduction}

The Aharonov-Bohm effect~\cite{0370-1301-62-1-303,PhysRev.115.485,PhysRev.123.1511} is a fundamental phenomenon of quantum interference related to the transmission of particles through a closed loop pierced by a magnetic flux. Besides it's fundamental significance for quantum theory, it's importance for applications in mesoscopic interferometric devices such as the electron Sagnac gyroscope~\cite{Toland2010923} is omnipresent. The effect was originally observed in metal rings in 1985~\cite{PhysRevLett.54.2696} and  
later also in carbon nanotubes~\cite{Bachtold:1999}. In graphene, an atomically thin two-dimensional carbon allotrope which was first isolated in 2004~\cite{Novoselov22102004}, the Aharonov-Bohm effect is expected to exhibit unusual behavior due to the peculiar electronic properties of this material, in which charge carriers at low energy behave effectively as massless Dirac fermions, giving rise to a number of (pseudo-)relativistic effects such as a Berry's phase $\pi$~\cite{Novoselov:2005,Zhang:2005} (for reviews on graphene see Refs.~\onlinecite{Geim:2007,RevModPhys.80.1337,RevModPhys.81.109}). In the following, we review recent developments on the Aharonov-Bohm effect in graphene nanostructures, first from the experimental side and, later on, we elaborate more on theoretical aspects.

\paragraph*{Experimental progress.}
The first experimental realization of a graphene ring structure was reported in 2008~\cite{PhysRevB.77.085413}. In this work, the authors investigate the Aharonov-Bohm oscillations in diffusive single-layer graphene as a function of the magnetic field which is applied perpendicular to the graphene plane in a two-terminal setup. They find clear magnetoconductance oscillations with the expected period corresponding to one magnetic flux quantum $\Phi_0 = h/e$ on top of a low-frequency background signal due to universal conductance fluctuations which are present in any disordered, phase-coherent mesoscopic device. Increasing the temperature $T$ gives rise to thermal averaging of the Aharonov-Bohm oscillations, and the authors find that the oscillation amplitudes decay as $T^{-1/2}$, as commonly observed in metal rings~\cite{doi:10.1080/00018738600101921}.

The authors further observe two unusual features in the recorded data. First, they find indications of a linear relationship between the oscillation amplitude and the overall ring conductance. Such a behavior has neither been observed in metal rings, nor in semiconductor heterostructures, with the exception of Ref.~\onlinecite{PhysRevB.75.115309}, where a similar effect is seen. The authors speculate that tunnel barriers which may be present in their device could be responsible for the observed behavior; however, a detailed theoretical analysis has yet to be done.

A second peculiar feature is the significant increase of the oscillation amplitude at strong magnetic fields close to the onset of the quantum Hall regime. This increase is strong enough to make the second harmonic---i.\,e.\ oscillations of period $\Phi_0 / 2 = h/2e$---visible in the frequency spectrum. Such a behavior was also observed by another group in subsequent experiments with smaller rings and higher visibility in a two-terminal as well as a four-terminal geometry and was attributed there to scattering on magnetic impurities~\cite{PSSB:PSSB200982284}---an explanation derived from corresponding observations in metallic rings that is not compatible with the observations made in Ref.~\onlinecite{PhysRevB.77.085413}, where, the authors instead speculate that the increase of the oscillation amplitude may be due to orbital effects originating from a potential asymmetry in the arms of the ring; however, this assumption could not be confirmed in subsequent numerical calculations~\cite{0268-1242-25-3-034003}.

\begin{figure}
	\begin{center}
		\includegraphics[width=\columnwidth]{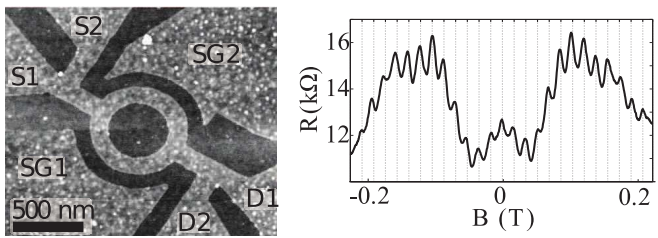}
	\end{center}
	\caption{(a) Scanning force micrograph and (b) magnetoresistance of the four-terminal (S1/2, D1/2) graphene ring structure with two side gates (SG1/2) investigated in Ref.~\onlinecite{1367-2630-12-4-043054} (Figures adapted from Ref.~\onlinecite{1367-2630-12-4-043054}).}
	\label{fig:exp}
\end{figure}

In Ref.~\onlinecite{PSSB:PSSB200982284}, the authors also introduce additional tunability into the graphene ring device by applying a side gate potential to one of the ring arms. In subsequent experiments~\cite{1367-2630-12-4-043054}, the same group systematically investigates the influence of such side gates in a four-terminal geometry (see Fig.~\ref{fig:exp}) in the diffusive regime and find phase shifts of the Aharonov-Bohm oscillations as a function of the gate voltage as well as phase jumps of $\pi$ at zero magnetic field---direct consequences of the electrostatic Aharonov-Bohm effect (which is more feasible in graphene than in metal rings due to the low screening of this material) as well as the generalized Onsager relations. %
The authors of Ref.~\onlinecite{PSSB:PSSB200982284} further speculate on the presence of edge disorder indicated by the fact that the various charge carrier trajectories around the ring, which can be derived from the frequency distribution of the magnetooscillations, do not cover the full area of the ring arms.

Further experiments on graphene ring structures include the local oxidation nanolithography using atomic force microscopy~\cite{weng:093107} and antidot arrays on epitaxial graphene films~\cite{shen:122102}; in the latter setup, universal conductance fluctuations are suppressed since the sample size exceeds the phase coherence length, while Aharonov-Bohm oscillations are still visible due to the small size of the antidots. As a graphene-specific feature, the authors also observe increased visibility of weak localization due to intervalley scattering on antidot edges. Shubnikov-de Haas and Aharonov-Bohm effects on thin graphite single crystals with columnar defects were investigated in Refs.~\onlinecite{springerlink:10.1134/S0021364009180167, 1742-6596-248-1-012001} in a four-probe measurement, and a significant contribution of surface Dirac fermions (``graphene on graphite'') as well as evidence supporting the theoretical prediction of edge states was found. %
Another multi-terminal measurement of Shubnikov-de Haas oscillations in monolayer graphene relating the Landau level separation between electrons and holes with the transport gap in the density of states is given in Ref.~\onlinecite{yoo:143112}; the poor visibility of the Aharonov-Bohm oscillations therein is attributed to a low phase coherence length.

\paragraph*{Theoretical progress.}
On the theoretical side, there is a great variety of topics which have been addressed, including the valley degree of freedom characteristic for graphene, particular device geometries and edge symmetries, resonant behavior and transistor applications, as well as the role of interactions. Common quantities expressing the influence of these aspects include electronic properties such as spectrum and persistent current, as well as transport properties such as conductance and noise.

In the pioneering work on the topic of Aharonov-Bohm rings made of graphene~\cite{PhysRevB.76.235404}, it was shown---both analytically using a circular ring in the weak intervalley scattering limit in the continuum model, as well as numerically using a hexagonal ring structure with zigzag terminated edges and strong intervalley scattering---that in the confined geometry of a graphene ring structure an applied magnetic flux gives rise to a lifting of the orbital degeneracy in a controllable fashion, which is manifest in conductance and persistent current, even in the absence of intervalley scattering. The magnitude of the lifting of valley degeneracy and it's dependence on details of the geometry of the ring have subsequently also been discussed in Ref.~\onlinecite{0953-8984-22-29-295503}. The effect was further observed in numerical magnetotransport simulations in the closed (or weakly coupled) circular ring geometry described by the tight-binding model~\cite{0268-1242-25-3-034003} for the case of smooth mass confinement, where intervalley scattering is suppressed. In this work, simulations have been done both for ballistic and diffusive regimes, and up to the quantum Hall regime where Aharonov-Bohm oscillations are found to be suppressed.  

Perfectly shaped ring devices with clean, well defined edges such as the hexagonal ring structure mentioned before~\cite{PhysRevB.76.235404} have been addressed in various studies, exploiting graphene-specific features of such ideal nanostructures. A detailed numerical study of the influence of shape and geometry, edge symmetries and corner structures on the electronic structure in the presence of a magnetic field reveals for example the edge state anticrossing and therefore gap opening due to the coupling of states localized at the inner and outer edges of the ring as well as the crucial role of corners in zigzag or armchair edge terminated rings~\cite{PhysRevB.79.125414}. For instance, the corners in an ideal hexagonal ring with zigzag edge termination in a magnetic field introduce intervalley scattering, as was considered in a supercell approach within the tight-binding model~\cite{B9NR00044E}; in this work, also a peculiar dependence of the spectrum as well as the persistent current on the even (resulting in semiconducting behavior) or odd (resulting in metallic behavior) character of the number of atoms across the ring arm width was found. Similar rings with metallic armchair termination have been considered in Refs.~\onlinecite{PhysRevB.80.165310,Fertig13122010}; in such systems, appropriately chosen corner junctions exhibit signatures of effective broken time reversal symmetry---caused by pseudomagnetic fields---at low energies, such as broken particle-hole symmetry or a gap in the spectrum that may be closed by the application of a real magnetic flux (The generation of pseudomagnetic fields in graphene rings under strain (or shear stress) has further been discussed in Ref.~\onlinecite{PhysRevB.84.115437}.). In Ref.~\onlinecite{Ma20101196}, the authors calculate the spectrum and the persistent current in ideal diamond shaped graphene ring structures either with zigzag or armchair edge termination within the tight-binding model; they also encounter the even-odd behavior mentioned before, and compare their results with previous work on hexagonal rings~\cite{B9NR00044E}.

Such ideal structures have also been found to be dominated by resonant behavior in the magnetotransport, for instance in Ref.~\onlinecite{0957-4484-21-18-185201}, where for small rectangular graphene nanorings with perfect edges resonant tunneling through quasi-bound states was observed rather than Aharonov-Bohm oscillations, which may be tuned by varying geometry, Fermi energy, or magnetic field. Resonant behavior was also observed in Ref.~\onlinecite{0268-1242-25-3-034003} as well as in in Ref.~\onlinecite{0957-4484-22-36-365201}, where it was proposed to utilize the electrostatic Aharonov-Bohm effect via side gates---such as already realized experimentally in Ref.~\onlinecite{1367-2630-12-4-043054}---for application in a quantum interference transistor with high on/off ratio made of a hexagonal graphene ring structure with perfect edges, where armchair edges were found to be preferable. In contrast, zigzag edges in the leads, acting as valley filters, and a circular graphene ring structure exhibiting an irregular boundary were considered in Ref.~\onlinecite{epub8537}, where resonant behavior was encountered as well; however, the main finding of this work was that for opposite valley polarization in the leads and appropriately sized rings exhibiting higher harmonics in the Fourier spectrum due to multiple turns around the ring region, the lowest harmonic is suppressed while higher harmonics are unaffected. A similar geometry, namely a graphene Aharonov-Bohm ring connected to valley filters which encircles a dislocation, was considered in the continuum model in Ref.~\onlinecite{PhysRevB.79.155111}, and decoherence properties were discussed. Possible applications as quantum interference transistors have also recently been discussed in Ref.~\onlinecite{Xu2012335} for disordered graphene rings, where ballistic rectification and negative differential resistance are observed in the $I$-$V$-characteristic; further, while for temperatures as large as $150$\,mK, phonon scattering is negligible, future work on the effect of electron-phonon interaction might be interesting.

Other aspects of interactions in graphene quantum rings have also been addressed in Ref.~\onlinecite{PhysRevB.78.193405} in the continuum model, with a focus on the interplay between valley polarization and Coulomb interaction, affecting the valley degeneracy. This influence was found to be accessible through the fractional nature of the periodicity of the Aharonov-Bohm oscillations in the persistent current as well as changes in the absorption spectrum. An analysis in the tight-binding model including electron-electron interactions further revealed the connection between electronic correlations and the spin polarization of the interacting ground state as function of ring size and number of electrons~\cite{Potasz:2009,PhysRevB.82.075425}. The interaction with the electromagnetic field has been discussed in Ref.~\onlinecite{PhysRevB.80.193407}, where it was found that, in a graphene ring threaded by a magnetic flux, excitations generated by electromagnetic pulses give rise to ``valley currents''---in analogy to spin currents.

The peculiar properties of graphene systems suggest to consider the Aharonov-Bohm effect also in more graphene-specific ring geometries and transport regimes. The electronic properties of monolayer as well as bilayer graphene rings in a magnetic field have been addressed in Refs.~\onlinecite{doi:10.1021/nl902302m,PhysRevB.81.045431,PhysRevB.82.119906}, either defining the ring geometry by tuning the band gap of bilayer graphene or employing a simplified zero-width ring geometry within the framework of the Dirac equation. In the magnetotransport within the (Andreev-)Corbino disk geometry in graphene, Landau level resonances, the suppression of conductance oscillations away from the charge neutrality point, and the crossover to the normal ballistic transport regime at large doping and weak fields have been encountered~\cite{PhysRevB.81.121404}. Analytical expressions for conductance and Fano factor in the magnetotransport of pseudodiffusive graphene rings, where transport at the Dirac point is dominated by evanescent modes, have been derived in Ref.~\onlinecite{0295-5075-89-1-17001}.

In the next sections, we will review our previous work on two graphene-specific effects in more detail. In Sec.~\ref{sec:klein}, Klein tunneling of Dirac fermions through a potential barrier in a graphene ring system similar to the experimental setup of Refs.~\onlinecite{PSSB:PSSB200982284,1367-2630-12-4-043054} is considered. A related effect---Andreev reflection at a graphene-superconductor interface in a mesoscopic ring device---is subsequently discussed in Sec.~\ref{sec:andreev}. In Sec.~\ref{sec:outlook} we will conclude and provide an outlook on potential future developments on the topic of the Aharonov-Bohm effect in graphene.

\section{Interplay of the Aharonov-Bohm effect and Klein tunneling in graphene}
\label{sec:klein}

\begin{figure}
	\centering
	\includegraphics[width=\columnwidth]{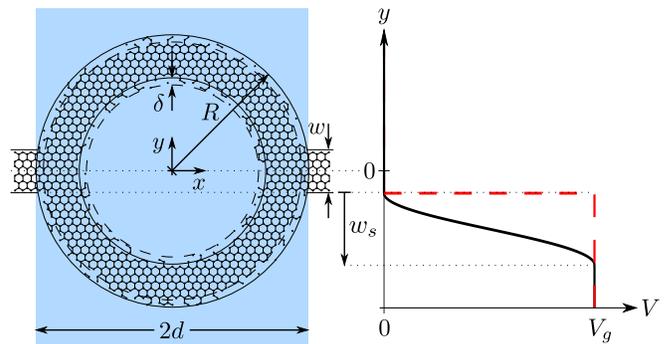}
	\caption{(Color online) Schematic of the graphene ring structure (left) and the $y$-dependence of the on-site gate potential $V$ (right) that is applied to the lattice sites on the lower arm of the ring. $V$ exhibits either a smooth (black solid line) or sharp (red dashed line) profile along $y$-direction while being constant along $x$-direction. The width of the arms of the ring is chosen equal to the width $w$ of the leads. The shaded area indicates the region of non-vanishing homogeneous magnetic field pointing out of plane. Different edge disorder configurations are realized by randomly removing sites within the two regions of width $\delta$ between dashed and solid circles (Figure adapted from Ref.~\onlinecite{PhysRevB.81.195441}).}
	\label{fig:setup}
\end{figure}

In this section, we review the numerical investigation of the effect of Klein tunneling~\cite{Katsnelson:2006,PhysRevB.74.041403} on the Aharonov-Bohm oscillations in a graphene ring on the basis of a tight-binding model with nearest-neighbor couplings. In order to introduce Klein tunneling into the system, an electrostatic potential can be applied to one of the arms of the ring (see Fig.~\ref{fig:setup} for a schematic), such that this arm together with the two adjacent leads form either a $nn'n$- or $npn$-junction ($n,n'$: conduction band transport, $p$: valence band transport). %
The former case corresponds to normal tunneling and the latter case to Klein tunneling. Then, the transmission properties strongly depend on the smoothness of the $pn$-interfaces. In particular, for sharp junctions the amplitude profile is symmetric around the charge neutrality point in the gated arm, whereas for smooth junctions the Aharonov-Bohm oscillations are strongly suppressed in the Klein tunneling as compared to the normal tunneling regime. Such a setup thus allows for a clear graphene-specific signature in Aharonov-Bohm measurements which seems to be readily observable. Its physical origin is the quantum interference of normal tunneling as well as Klein tunneling trajectories through the two arms of the ring. The work presented in this section was originally published in Ref.~\onlinecite{PhysRevB.81.195441}.

\subsection{Model}

The calculation is based on the usual tight-binding Hamiltonian for graphene
\begin{equation}
	H = \sum_i{ V_i \left| i \right\rangle \left\langle i \right| } + \sum_{ \left\langle i, j \right\rangle }{ \tau_{ij} \left| i \right\rangle \left\langle j \right| },
	\label{for:TBM}
\end{equation}
where the second sum runs over nearest-neighbors and $V_i = V(\mathbf{r}_i)$ is a position-dependent on-site potential, taking the origin of coordinates at the center of the ring. The graphene hopping integral $\tau_0 \sim 2.7$\,eV picks up a Peierls phase in the presence of a magnetic field, yielding for the nearest-neighbor coupling element the expression
\begin{equation}
	\tau_{ij} = -\tau_0 \, \exp{ \left( \frac{ 2 \pi \mathrm{i} }{ \Phi_0 } \int_{ \mathbf{r}_i }^{ \mathbf{r}_j }{ \mathbf{A}(\mathbf{r}) \, \mathrm{d}\mathbf{r} } \right) },
	\label{for:Peierls}
\end{equation}
where the line integral is taken along the straight path between sites $i$ and $j$. $\Phi_0 = h / e$ is the magnetic flux quantum, and
\begin{equation}
	\mathbf{A}(\mathbf{r}) = -B \, y \, \theta(d - |x|) \, \mathbf{\hat{e}}_x
	\label{for:vecPot}
\end{equation}
with $d = \sqrt{R^2 - w^2/4}$ is the vector potential giving rise to a homogeneous magentic field
\begin{equation}
	\mathbf{B}(\mathbf{r}) = \boldsymbol{\nabla} \times \mathbf{A}(\mathbf{r}) = B \, \theta(d - |x|) \, \mathbf{\hat{e}}_z.
\end{equation}

The system under consideration is a ring-shaped structure cut out of a graphene sheet, which is attached to two crystalline leads also modeled using the graphene lattice structure (see Fig.~\ref{fig:setup}). Besides the magnetic field, the structure is also subject to a gate electrode potential $V_g$ located on top of the lower arm of the ring. The smoothness of the potential interface is controlled via the smoothing width $w_s$ measured from the lower edges of the leads:
\begin{align*}
	V = 0 \quad &\mathrm{for} \quad y \geq -w / 2,\\
	V = V_g	 \quad &\mathrm{for} \quad y \leq -w / 2 - w_s,\\
	0 < V < V_g \quad &\mathrm{otherwise}.
\end{align*}
In the presented simulations, a cosine-shaped smoothing profile is used and $0 \leq w_s \leq R - 3w/2$.

\begin{figure}
	\centering
	\includegraphics[width=\columnwidth]{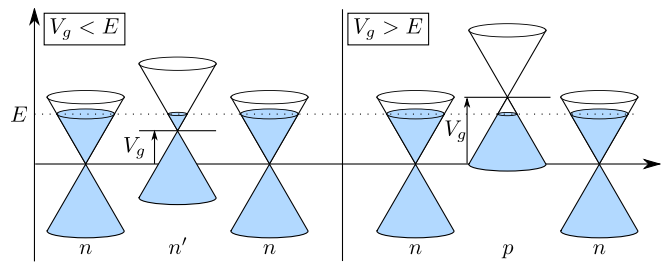}
	\caption{(Color online) Schematic of the influence of the potential profile introduced by $V_g$ on the spectrum of the lower arm of the ring. The left hand side shows the normal tunneling case ($nn'n$-junction) and the right hand side the Klein tunneling case ($npn$-junction) (Figure adapted from Ref.~\onlinecite{PhysRevB.81.195441}).}
	\label{fig:junctionDirac}
\end{figure}

For a Fermi energy $E > 0$, together with the adjacent leads this lower arm forms either a $nn'n$- or $npn$-junction for $V_g < E$ and $V_g > E$, respectively (see Fig.~\ref{fig:junctionDirac} for a schematic; the Fermi energy is measured relative to the charge neutrality point in the leads). %
Note that the setup exhibits a flat potential profile for trajectories along the upper ring arm, i.\,e.\ a $nnn$-junction, since there is no gate potential applied. This enables a rather large transmission through the ring even when the lower ring arm is tuned towards the Dirac point, since transport through the upper arm always takes place at an energy distance $E$ away from the charge neutrality point.

Transport calculations are done in the Landauer-B\"uttiker formalism for elastic transport at zero temperature assuming complete phase-coherence using a variant of the recursive Green's function technique~\cite{PhysRevLett.47.882,0022-3719-14-3-007,springerlink:10.1007/BF01328846} for the coupling of the surface Green's functions of the leads, as described in Ref.~\onlinecite{PhysRevB.81.195441}; the linear conductance of the system is expressed through the transmission function $\mathrm{Tr}(t^\dagger t)$, where $t$ is the $N \times N$ transmission matrix between the two leads, i.\,e.\ the lower left block of the unitary scattering matrix $S$,
\begin{equation}
	S = \left( 
	\begin{array}{cc}
		r & t'\\
		t & r'
	\end{array}
	\right),
\end{equation}
which is obtained from the Green's function of the coupled system via a Fisher-Lee relation. $N$ is the number of propagating modes in the leads.

\subsection{Results}

\begin{figure}
	\centering
	\includegraphics[width=\columnwidth]{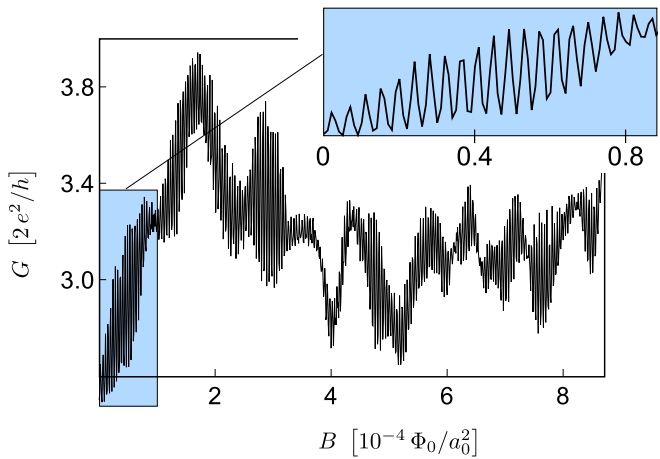}
	\caption{(Color online) Magnetoconductance of a ring with $R / a_0 = 300$, $w / a_0 = 60$ at energy $E / \tau_0 = 0.5$ and zero gate voltage, showing clear Aharonov-Bohm oscillations on top of a background due to universal conductance fluctuations (Figure adapted from Ref.~\onlinecite{PhysRevB.81.195441}).}
	\label{fig:magnetoconductance}
\end{figure}

\begin{figure}
	\centering
	\includegraphics[width=\columnwidth]{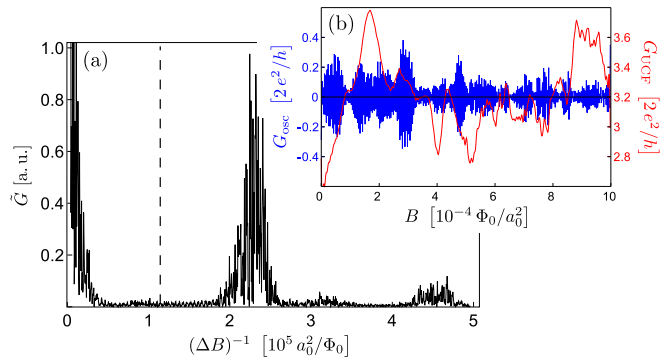}
	\caption{(Color online) (a) Frequency spectrum corresponding to Fig.~\ref{fig:magnetoconductance}, obtained from the Fourier transform $\tilde{G}$ of the magnetoconductance $G$. Besides the low frequency background and the fundamental oscillation component, the second harmonic is also slightly visible in the spectrum. The dashed line indicates the frequency limit of the high pass frequency filter used for background subtraction (Figure adapted from Ref.~\onlinecite{PhysRevB.81.195441}). (b) UCF background signal (red) and magnetooscillations (blue) corresponding to Fig.~\ref{fig:magnetoconductance} after background subtraction.}
	\label{fig:fourierSpectrum}
\end{figure}

In the following, we present transmission properties for a ring with $R / a_0 = 300$ and $w / a_0 = 60$, $a_0 = 0.142$\,nm being the nearest-neighbor distance in graphene, in terms of the linear conductance $G = 2e^2/h \cdot \mathrm{Tr}(t^\dagger t)$, where the factor 2 accounts for spin degeneracy. Edge disorder is applied to the ring by randomly removing sites within a width $\delta$ from the inner and outer edges of the ring, respectively (see Fig.~\ref{fig:setup}). We choose $\delta / a_0 = 1.5$ in order to keep the edge of the ring as smooth as possible while still allowing for different edge disorder configurations. Fermi energy $E$ and gate potential $V_g$ are chosen such that transport always takes place in between the van Hove singularities located at $E = \pm \tau_0$ where the density of states diverges in the tight binding model of graphene, $0 < E < \tau_0$, $0 \leq V_g \leq 2 \, E$. In Fig.~\ref{fig:magnetoconductance}, we plot the magnetoconductance at Fermi energy $E / \tau_0 = 0.5$ and zero gate voltage ($V_g = 0$) for a particular ring realization, showing pronounced Aharonov-Bohm oscillations on top of a low frequency background. %
The background signal results from universal conductance fluctuations (UCF) which are typical for phase-coherent mesoscopic devices~\cite{datta2005electronic}. The behavior is in agreement with the observations made in Ref.~\onlinecite{0268-1242-25-3-034003}, where the authors investigate an even wider magnetic field range up to the quantum Hall regime. In Fig.~\ref{fig:fourierSpectrum}, we also show the corresponding frequency spectrum obtained from a Fourier transform of the magnetoconductance signal up to $B = 10^{-3} \, \Phi_0 / a_0^2$, as well as the UCF background signal and the magnetooscillations after background removal by means of a high pass frequency filter. %
The contributions to the Aharonov-Bohm oscillations are centered around $(\Delta B \, a_0^2 \, e / h)^{-1} \sim 2.3 \cdot 10^5$. Using $\tilde{R}^2 \, \pi \cdot \Delta B = h / e$, this frequency corresponds to a mean radius $\tilde{R} / a_0 \sim 270$ of interfering electron trajectories, which perfectly lies within the boundaries of the ring.

By applying a gate voltage $V_g > 0$ to one of the ring arms, the magnitude of the Aharonov-Bohm oscillations may be modified. A convenient measure of the oscillation magnitude is the root mean square (RMS) amplitude of the signal. Prior to the RMS analysis, the UCF background has to be removed from the signal. This is achieved by applying a high pass frequency filter to the Fourier transform of the magnetoconductance data, as indicated in Fig.~\ref{fig:fourierSpectrum}. The retained, unbiased signal is squared, and the root of the average over the squared signal is defined as the RMS amplitude $\Delta G_{\text{RMS}}$.

\begin{figure}
	\centering
	\includegraphics[width=\columnwidth]{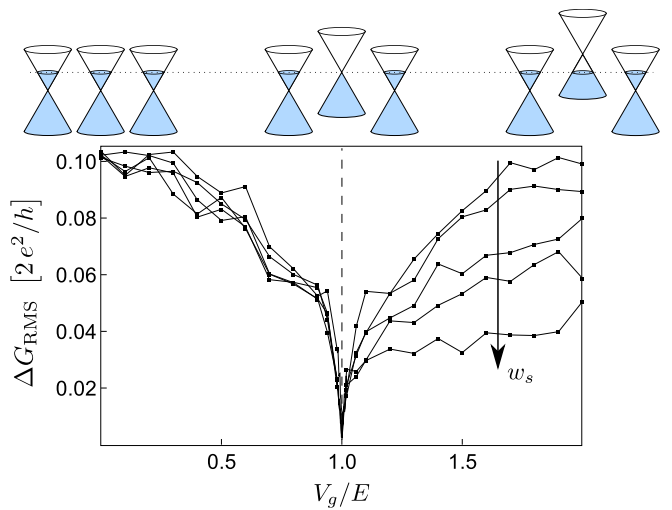}
	\caption{(Color online) RMS analysis for the setup used in Fig.~\ref{fig:magnetoconductance} for different smoothing widths $w_s / a_0 \in \{ 0, 21, 52.5, 105, 210 \}$. Each data point results from an average over five realizations of edge disorder. The corresponding standard deviations lie between $0.005 \cdot 2e^2/h$ and $0.015 \cdot 2e^2/h$ but are suppressed for better visibility. For better clarity, the spectrum schematics (see Fig.~\ref{fig:junctionDirac}) are also included (Figure adapted from Ref.~\onlinecite{PhysRevB.81.195441}).}
	\label{fig:rmsPlot}
\end{figure}

In Fig.~\ref{fig:rmsPlot} we show the dependence of the RMS oscillation amplitude $\Delta G_{\text{RMS}}$ on the gate voltage $V_g$ for different smoothing widths $w_s$ (see Fig.~\ref{fig:setup}) at energy $E / \tau_0 = 0.5$, where the average is taken over the range $0 \leq B a_0^2 / \Phi_0 \leq 10^{-3}$. %
Increasing the gate voltage from zero towards the neutrality point $V_g = E$ not only leads to increased potential scattering but also to a reduction in the number of accessible propagating states in the lower arm of the ring. As can be seen in Fig.~\ref{fig:rmsPlot}, the oscillation amplitude diminishes and reaches a minimum value at the neutrality point. Note that, since the transmission through the upper ring arm is not at all affected by a gate potential, the overall conductance itself is only slightly changed to fluctuate around $2.5 \cdot 2e^2/h$ (see Fig.~7 in Ref.~\onlinecite{PhysRevB.81.195441}) as compared to values around $3.4 \cdot 2e^2/h$ in the case of zero gate potential on the lower ring arm (see Fig.~\ref{fig:magnetoconductance}).

For $V_g < E$, the decay of the RMS amplitude towards the neutrality point does not depend on the details of the gate potential interface. However, in the regime of Klein tunneling, $V_g > E$, the oscillation behavior strongly depends on the smoothness of the gate potential. In case of a smooth potential, the partial waves in the lower arm have to tunnel through a finite region of low density of states, where $V \sim E$ (see Fig.~\ref{fig:setup}), in order to interfere with the partial waves traversing the upper arm. The lower arm becomes increasingly penetrable as this region gets narrower, until it gets transparent in case of a sharp potential. This reflects the usual behavior of Klein tunneling phenomena, where the probability for tunneling through a $pn$-junction depends on the smoothness of the $pn$-interface~\cite{RevModPhys.80.1337,PhysRevB.74.041403}.

The described behavior of the RMS amplitude is robust over the whole energy range under consideration, except for an increasing uncertainty at lower values for the Fermi energy. Although all results are presented for zigzag boundary conditions in the leads, the effects are independent of a change of orientation of the graphene lattice to armchair boundaries in the leads.

Klein tunneling in graphene thus exhibits clear signatures in the Aharonov-Bohm oscillations observed in mesoscopic rings. In the next section, we will show how another graphene-specific effect, namely specular Andreev reflection at a graphene-superconductor interface~\cite{PhysRevLett.97.067007}, can be identified in such nanostructures.

\section{How to distinguish specular from retro Andreev reflection in graphene rings}
\label{sec:andreev}

In this section, we review numerical transport calculations of Andreev reflection in a graphene ring system threaded by a magnetic flux and attached to one normal conducting and one superconducting lead. To this end, the Bogoliubov-de Gennes equation for the tight binding model using the recursive Green's functions technique is solved within the Landauer-B\"uttiker framework for elastic transport. By tuning chemical potential and bias voltage, it is possible to switch between regimes where electron and hole originate from the same band (\emph{retro configuration}) or from different bands (\emph{specular configuration}) of the graphene dispersion, respectively. Andreev reflection is known to be closely related to the effect of Klein tunneling discussed in the previous section.~\cite{PhysRevB.77.075409} However, different aspects of Klein tunneling have become experimentally accessible in the last years~\cite{PhysRevLett.102.026807,Young:2009}, whereas specular Andreev reflection has not been observed to date, although there exist a number of proposals for the experimental control~\cite{PhysRevLett.103.167003} and detection~\cite{PhysRevLett.97.067007,PhysRevB.75.045417,PhysRevB.83.235403} of this process. (For a review on both effects, see Ref.~\onlinecite{RevModPhys.80.1337}.) Here, we review a novel approach concerning the identification of specular Andreev reflection, distinguishing it from conventional retro reflection, and discuss the advantages over previous works in the field. We find that the dominant contributions to the Aharonov-Bohm oscillations in the subgap transport are of period $h/2e$ in retro configuration, whereas in specular configuration they are of period $h/e$. This result confirms the predictions obtained from a qualitative analysis of interfering scattering paths, and since it is robust against disorder and moderate changes of the system, it provides a clear signature to distinguish both types of Andreev reflection processes in graphene. The work presented in this section was originally published in Ref.~\onlinecite{2011arXiv1110.4383S}.

\subsection{Scattering path analysis}

\begin{figure}
	\centering
	\includegraphics[width=\columnwidth]{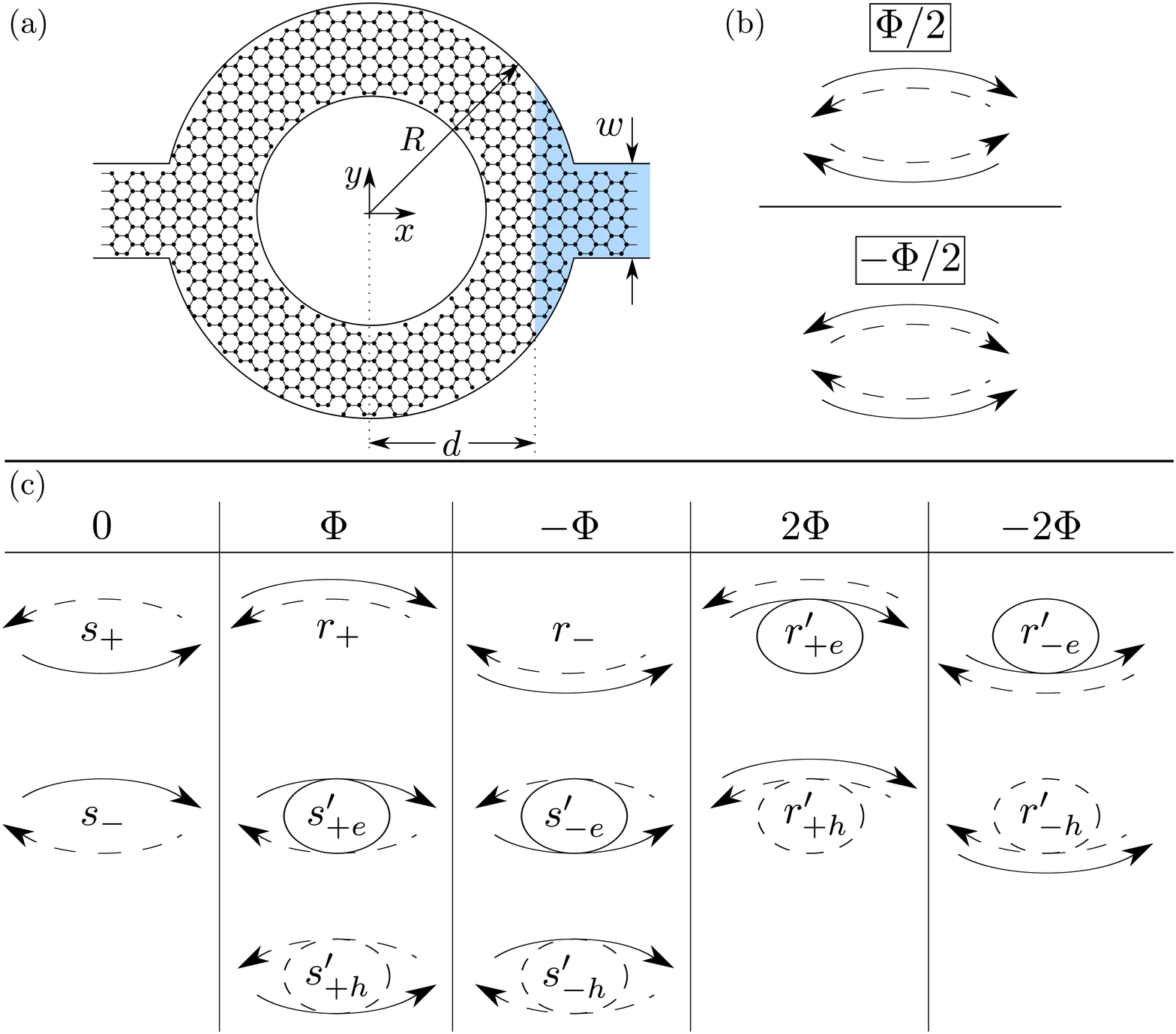}
	\caption{(Color online) (a) Device geometry showing a graphene ring structure that is penetrated by a magnetic flux $\Phi$ measured in units of the flux quantum $\Phi_0$. At the interface with the superconductor (shaded region), electron-hole conversion may occur. (b) The gauge is chosen such that each of the eight individual electron (solid lines) and hole (dashed lines) paths picks up a phase $\pm \Phi/2$ as indicated. (c) Scattering paths for electrons injected from and holes leaving through the left normal conducting lead; only zeroth and first order contributions are included, i.\,e.\ terms containing a single electron-hole conversion process and at most one additional round-trip of the electron or the hole. The paths are categorized according to the total phase that is picked up, and each path is associated with a corresponding amplitude, where first order amplitudes are indicated by a prime (Figure adapted from Ref.~\onlinecite{2011arXiv1110.4383S}).}
	\label{fig:paths}
\end{figure}

Our approach is based on the observation, that in general, the probability for an incident electron to be reflected as a hole is less than one. This allows for effects typical for phase-coherent mesoscopic devices, like universal conductance fluctuations or Aharonov-Bohm oscillations~\cite{PhysRev.115.485} in the magnetoconductance. While in normal metals, the fundamental period of these oscillations is given by the flux quantum $\Phi_0 = h/e$, it is half the value for Andreev (retro) reflection in conventional metals, due to the charge $2e$ of a Cooper pair. However, this is not true anymore in the case of specular Andreev reflection, therefore providing a criterion to distinguish between specular and retro reflection. In order to show this, we consider the phases due to the magnetic flux that are picked up by the various scattering paths. In this analysis, we restrict ourselves to the contributions up to first order in the sense that we take processes into account that involve only a single electron-hole conversion process, and that contain at most one additional round-trip of electron or hole, respectively; higher order contributions connected with additional round-trips are often times negligible~\cite{PhysRevB.81.195441,PhysRevB.77.085413}. The corresponding 
paths are summarized in Fig.~\ref{fig:paths}. %
In order to obtain the magnetoconductance for the two types of Andreev reflection, we sum up the amplitudes as defined in Fig.~\ref{fig:paths} for the various paths coherently:
\begin{align*}
	R_s(\Phi) \cong& \left| 
	s_+ + s_- + s_+' \, \E^{\I \Phi} + s_-' \, \E^{-\I \Phi}
	\right|^2 \nonumber \\
	R_r(\Phi) \cong& \left| 
	r_+ \, \E^{\I \Phi} + r_- \, \E^{-\I \Phi} + 
	r_+' \, \E^{2 \I \Phi} + r_-' \, \E^{-2 \I \Phi}
	\right|^2
\end{align*}
where $s_\pm' = s_{\pm e}' + s_{\pm h}'$, $r_\pm' = r_{\pm e}' + r_{\pm h}'$, and $\Phi$ is the magnetic flux measured in units of the flux quantum $\Phi_0$. 
Assuming $|s| \gg |s'|$ for any zeroth- and first-order amplitudes, respectively, we obtain
\begin{equation}
\label{for:Ts}
	R_s(\Phi) \cong R_s^0 + 
	2 \, \mathrm{Re} \left[(s_+' \, s_0^* + s_0 \, s_-'^*)  \, \E^{\I \Phi}\right] + 
	\mathcal{O}[(s')^2],
\end{equation}
where $s_0 = s_+ + s_-$ and $R_s^0$ contains contributions that are constant with respect to $\Phi$. Therefore, in the case of specular reflection, oscillations of period $h/e$ are dominant. In contrast, in the case of retro reflection, contributions of period $h/2e$ are dominant, as expected:
\begin{equation}
\label{for:Tr}
	R_r(\Phi) \cong R_r^0 +
	2 \, \mathrm{Re} \left[r_+ \, r_-^* \, \E^{2 \I \Phi}	\right] + \mathcal{O}[r \, r', \ (r')^2],
\end{equation}
where again $R_r^0$ contains $\Phi$-independent terms and we assume $|r| \gg |r'|$ for any zeroth- and first-order amplitudes, respectively.

\subsection{Numerical model}

In order to test the previous analysis on the basis of a microscopic model, we implement the Bogoliubov-de Gennes Hamiltonian~\cite{gennes1999superconductivity}
\begin{equation}
\label{for:hamBdG}
\mathcal{H} = \left(
	\begin{array}{cc}
		H	 - E_F			& \boldsymbol{\Delta}\\
		\boldsymbol{\Delta}				& E_F - H^*
	\end{array}
	\right)
\end{equation}
within the tight binding formalism of graphene, Eq.~\eqref{for:TBM}. In this numerical calculation, all higher order contributions beyond the ones discussed in the previous subsection are also taken into account. In Eq.~\eqref{for:hamBdG}, we assume $\Delta_i = \Delta(\mathbf{r}_i) \in \mathbb{R}$ for the superconducting order parameter $\boldsymbol{\Delta} = \sum_i \Delta_i \ket{i} \bra{i}$. The presence of a magnetic field is captured by a Peierls phase in the hopping matrix element, Eq.~\eqref{for:Peierls}.

The structure of the graphene device under consideration is schematically shown in Fig.~\ref{fig:paths}. The two semi-infinite leads also exhibit the graphene lattice structure; superconductivity is induced into the right lead due to the proximity effect of a superconducting electrode on top of the graphene. We choose to orient the leads to exhibit armchair edges and later comment on the reason for this particular choice. The whole ring is penetrated by a uniform perpendicular magnetic field of strength $B$, described by the vector potential~\eqref{for:vecPot}. The origin of coordinates is taken at the center of the ring.

\begin{figure}
	\centering
	\includegraphics[width=\columnwidth]{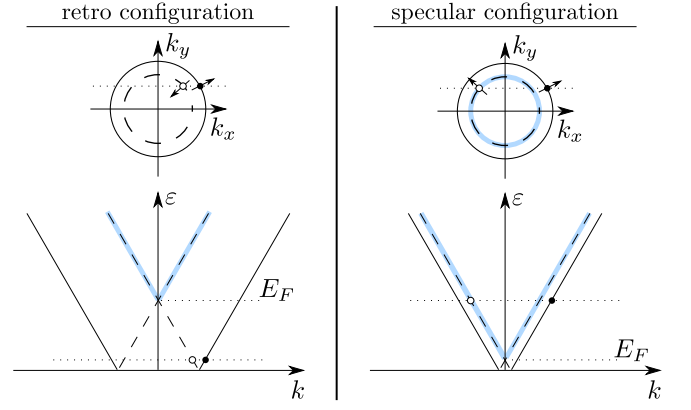}
	\caption{(Color online) Schematics of the (radially symmetric) excitation spectrum (lower panel) and surfaces of constant excitation energy in $\mathbf{k}$-space (upper panel) in the cases $E_F > \eps > 0$ (\emph{retro configuration}) and $0 < E_F < \eps$ (\emph{specular configuration}). Solid and dashed lines indicate electron- and hole-like states, respectively, (hole) states originating from the valence band are shaded. The small arrows in the upper panel indicate the direction of propagation of the corresponding states. Electron-hole excitations are drawn assuming conservation of $k_y$ at the NS interface (Figure adapted from Ref.~\onlinecite{2011arXiv1110.4383S}).}
	\label{fig:dispersion}
\end{figure}

In order to fulfill the mean-field requirement of superconductivity, which demands the superconducting coherence length $\xi = \hbar v_F / \Delta$ to be large compared to the wavelength $\lambda_S$ in the superconducting region~\cite{PhysRevLett.97.067007}, we introduce additional doping into the superconducting region by applying a gate potential $V_i = V_g \, \theta(x_i - d)$. Which type of Andreev reflection occurs at the NS interface is then determined by the excitation energy $\eps$ (i.\,e.\ the eigenvalues of Eq.~\eqref{for:hamBdG}) and the Fermi energy $E_F$, as shown in Fig.~\ref{fig:dispersion}. %
In \emph{retro configuration}, $E_F > \eps > 0$, where $v_y^{(h)} \cdot v_y^{(e)} < 0$ for the $y$-components of the electron and hole velocities, both electron and hole traverse the same arm of the ring. In \emph{specular configuration}, $0 < E_F < \eps$, the hole is reflected back through the other arm of the ring, since $v_y^{(h)} \cdot v_y^{(e)} > 0$. In the following, we choose $|V_g| \gg E_F$, justifying the adoption of the step-function model for the superconducting order parameter, $\Delta_i = \Delta \, \theta(x_i - d)$~\cite{PhysRevLett.97.067007}.

In order to compare retro (r) and specular (s) configurations, we will choose $\eps^{(r)} = E_F^{(s)}$ and $\eps^{(s)} = E_F^{(r)}$ since then the states in both configurations exhibit the same wavelength and there is the same number of propagating modes. We further choose $\eps^{(r)}, E_F^{(s)} \ll \eps^{(s)}, E_F^{(r)}$ so that for nearly each value of $k_y$, there exist electron-hole scattering channels.

Again, the transport properties of the system are obtained from the scattering matrix $S$ that is calculated in the framework of the Landauer-B\"uttiker formalism using the recursive Green's function technique as in the previous section. In the framework of the Bogoliubov-de Gennes Hamiltonian, Green's function and scattering matrix are parameterized by the eigenvalues $\eps$ of the Hamiltonian \eqref{for:hamBdG}.

In the following, we will concentrate on the regime $\eps < \Delta$, in which there are no propagating modes in the superconducting lead, so that electrons injected from the normal conducting lead are reflected back either as electron  (e) or hole (h). The scattering matrix thus has the structure
\begin{equation*}
	S = \left(
	\begin{array}{cc}
		r_{ee}	& r_{eh} \\
		r_{he}	& r_{hh}
	\end{array}
	\right)
\end{equation*}
from which the differential conductance for the Andreev processes is given by
\begin{equation*}
	\frac{dI}{dV} = \frac{4e^2}{h} \cdot \mathrm{Tr}(r_{he}^\dagger \, r_{he})
\end{equation*}
where the factor 4 accounts for spin degeneracy and the quantization of charge in units of $2e$.

\subsection{Results}

\begin{figure}
\centering
\includegraphics[width=\columnwidth]{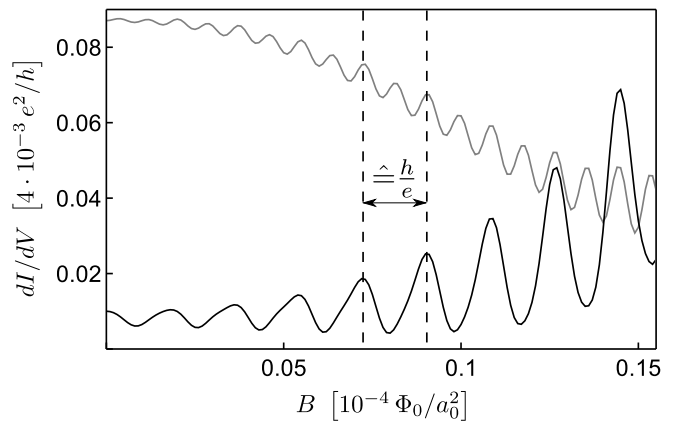}
\caption{Differential magnetoconductance for specular (black) and retro (gray) configuration for $E_F^{(r)} = 0.025\,\tau_0 = \eps^{(s)}$, $E_F^{(s)} = 0.001\,\tau_0 = \eps^{(r)}$, corresponding to 8 modes in the normal conducting lead, including all degeneracies (spin, valley, electron/hole). The high doping in the superconducting lead is chosen such that $E_F - V_g = 0.5\,\tau_0$ in both cases. Other parameter values are provided in the main text. The period of the dominant oscillation is $B_0^{(s)} \approx 1.8 \cdot 10^{-6} a_0^{-2} h/e$ in specular configuration and $B_0^{(r)} \approx 8.8 \cdot 10^{-7} a_0^{-2} h/e \approx 0.5 \, B_0^{(s)}$ in retro configuration. The weak beating pattern in retro configuration and the asymmetry in specular configuration arise due to minor contributions of contrary frequencies (Figure adapted from Ref.~\onlinecite{2011arXiv1110.4383S}).}
\label{fig:results}
\end{figure}

In Fig.~\ref{fig:results}, we show the calculated transmission for a ring of width $w = 87 \, \sqrt{3} \, a_0$ and outer radius $R = 500 \, a_0$, where $a_0$ is the distance between nearest neighbors. %
The transmission function exhibits Aharonov-Bohm oscillations on top of a low frequency background which is due to universal conductance fluctuations. The position of the NS interface is given by $d = 400 \, a_0$. The chosen dimensions of the ring are large enough to exclude finite-size effects 
while still being numerically manageable. For the superconducting order parameter, we choose a value of $\Delta = 0.03 \, \tau_0 \approx 80\,$meV, which may appear unrealistic at first sight, considered the fact that typical values are up to a few meV. However, by making this choice we scale the value of the superconducting order parameter according to the scale of the system size, such that the dimensionless factor $\Delta R / \hbar v_F$ stays of same order of magnitude, compared with values realized in experiments~\cite{PhysRevB.77.085413,1367-2630-12-4-043054}. Thus, for a realistic system size of $R \sim 10^{-6}$m, our choice of $\Delta$ would correspond to a value of a few meV for the superconducting gap. Note that due to these low energy scales and the rather large spacing of modes resulting from the narrow geometry of the electron waveguides in such a ring structure, in specular configuration only the regime of a low number of modes is accessible. Also note that due to strong electron backscattering at the front of the hole and at the rough edges of the ring, the average value of the differential conductance is much less than a conductance quantum, $e^2/h$.

The average radius $\bar{r}$ of the scattering path is calculated according to $\bar{r}^2 \pi B_0 = h / n e$, where $n = 1$ ($n = 2$) in specular (retro) configuration and $B_0$ is the (dominant) period of the oscillation. Evaluating the period of the oscillations shown in Fig.~\ref{fig:results}, we obtain $\bar{r}^{(s)} \approx 420\,a_0$ in specular configuration and $\bar{r}^{(r)} \approx 425\,a_0$ in retro configuration. The obtained values lie well within the inner and outer radius of the ring and close to the arithmetic mean $R - w / 2 \approx 425\,a_0$, therefore confirming the predictions obtained from Eqs.~\eqref{for:Ts} and~\eqref{for:Tr}. Minor contributions of period $h/e$ in retro configuration and $h/2e$ in specular configuration visible in Fig.~\ref{fig:results} may arise due to terms neglected in Eqs.~\eqref{for:Ts} and ~\eqref{for:Tr}, scattering off the sharp boundaries of the ring structure, and the fact that for the electron-hole conversion at the NS interface $k_y$ is not strictly conserved.

\begin{figure}
\centering
\includegraphics[width=\columnwidth]{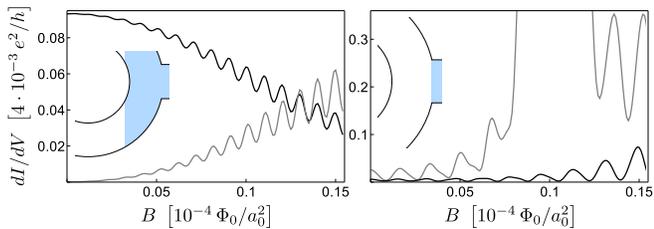}
\caption{(Color online) Breakdown of the $h/e$ vs. $h/2e$ signature for shifted positions of the NS interface, as explained in the text. Other parameters and color coding 
	are chosen as in Fig.~\ref{fig:results}. For $d = 340\,a_0$ (left), in specular configuration one observes oscillations of period $h/2e$ as in retro configuration. For $d = 490\,a_0$ (right), contributions of specularly reflected holes in retro configuration become important, leading to the observation of additional $h/e$-oscillations. The value of the superconducting coherence length is $\xi = 50\,a_0$ (Figure adapted from Ref.~\onlinecite{2011arXiv1110.4383S}).}
\label{fig:breakdown}
\end{figure}

Another strong evidence that supports our interpretation of the two different periods is the breakdown of this particular signature that is observed for a shift of the position of the NS interface on the scale of the width of the ring. Indeed, while in Ref.~\onlinecite{PhysRevB.81.174523}---where a three-terminal graphene junction is analyzed---the exact position of the NS interface has no effect, it matters here; the reason is that $\xi$ is comparable or even less than the system size, while in Ref.~\onlinecite{PhysRevB.81.174523} the superconducting coherence length greatly exceeds the system dimensions. If the interface is too close to the hole region (see Fig.~\ref{fig:breakdown}\,(a) inset), then specularly reflected holes are forced to traverse the same arm as the incoming electron. %
In this case, one should observe $h/2e$ oscillations in specular configuration. In contrast, if the interface is too far from the hole (see Fig.~\ref{fig:breakdown}\,(b) inset), holes may significantly be reflected through the other arm, e.\,g.\ due to increased scattering at the ring boundaries. This would manifest itself in the observation of $h/e$ oscillations in addition to the $h/2e$ oscillations in retro configuration. This behavior is confirmed in the observed magnetooscillations, as shown in Fig.~\ref{fig:breakdown}.

\begin{figure}
	\begin{center}
		\includegraphics[width=\columnwidth]{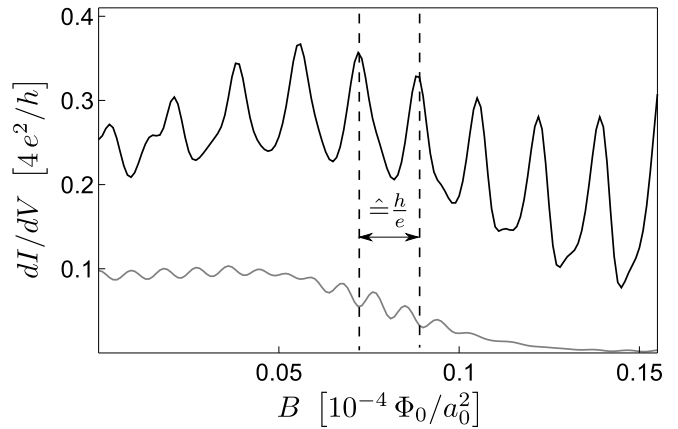}
	\end{center}
	\caption{Differential magnetoconductance for $d=430\,a_0$, $E_F^{(r)} = 0.029\,\tau_0 = \varepsilon^{(s)}$, $E_F^{(s)} = 0.0001 \, \tau_0 = \varepsilon^{(r)}$, corresponding to twice the number of propagating modes, compared to Fig.~\ref{fig:results}. Other parameters and color coding are chosen as in Fig.~\ref{fig:results}. Note that the magnetoconductance signal is strongly enhanced compared to Fig.~\ref{fig:results}.}
	\label{fig:moreModes}
\end{figure}

\begin{figure}
	\centering
	\includegraphics[width=\columnwidth]{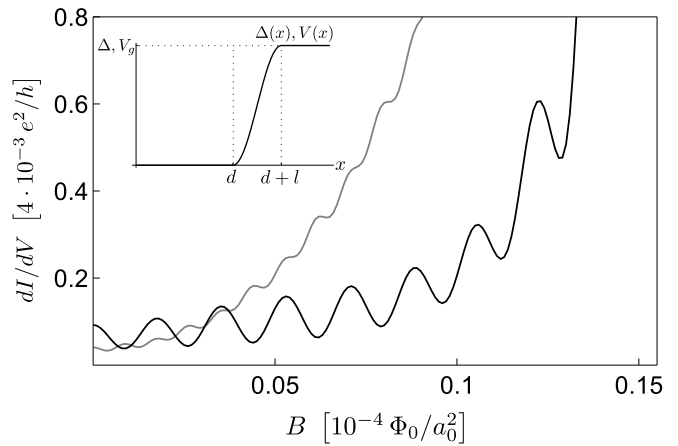}
	\caption{Magnetoconductance of the system used in Fig.~\ref{fig:results} with a smooth potential profile (inset) with $l = 90\,a_0$ and bulk disorder of strength $\sigma = 0.01\,\tau_0$ as explained in the text. The $h/e$ vs. $h/2e$ signature still persists. The color coding is the same as in Fig.~\ref{fig:results} (Figure adapted from Ref.~\onlinecite{2011arXiv1110.4383S}).}
	\label{fig:disorder}
\end{figure}

Apart from that, the $h/e$ vs. $h/2e$ signature proves to be very robust against moderate changes to the length and energy scales in the system, such as the extent of the magnetic field or the ratio of Fermi wavelength and the width of the NS interface. We also tested that the signature persists when more propagating modes are present in the lead, leading to values of the average conductance which are much larger compared to the few-mode situation shown in Fig.~\ref{fig:results} (see Fig.~\ref{fig:moreModes}).
Additionally, the signature is hardly affected by bulk disorder, which is a major advantage of our setup. In Fig.~\ref{fig:disorder}, we show the magnetoconductance of the system used in Fig.~\ref{fig:results} with a particular random short-range disorder configuration, which is realized by applying an uncorrelated, random on-site potential of Gaussian distribution with zero mean and width $\sigma = 0.01\,\tau_0$ to each site. %
In addition, the NS interface has been smeared out over a distance $l = 90\,a_0$ in this case. The robustness of the effect can be explained from the topological nature of the signature: since all microscopic scattering paths  can be classified into just two groups---yielding $h/e$- or $h/2e$-oscillations, respectively---according to which arm is traversed by the quasiparticles, impurity scattering and the resulting deflection of quasiparticles has no adverse effect as long as scattering \textit{between} the groups is weak, while scattering \textit{within} one group may be arbitrarily strong. In addition, note that while our description of transport via the scattering matrix assumes complete phase coherence, a signature that distinguishes retro from specular Andreev reflection is assumed to persist also in the case of a finite phase coherence length. More specifically, if the phase coherence length is on the order of the ring circumference, first-order amplitudes in Eqs.~\eqref{for:Ts} and \eqref{for:Tr} may be neglected. Then, retroreflection would still manifest itself in $h/2e$-oscillations, while there would be no oscillations at all in the case of specular reflection.

Finally, we add a remark concerning the choice of armchair boundary conditions in the leads employed in the analysis in this section. In a tight binding implementation of graphene, there are two simple choices for the orientation of the leads. Often, zigzag edges are considered to represent a generic boundary condition for graphene ribbons~\cite{PhysRevB.77.085423}. In this case, edge states are present in the system that modify the simple picture provided in Fig.~\ref{fig:dispersion} by adding additional scattering channels between bulk and edge states while removing certain scattering channels between bulk states due to the conservation of the so-called pseudoparity symmetry that acts like a selection rule~\cite{PhysRevB.79.115131}. In the realistic limit of metal leads providing a large number of propagating bulk modes, this effect should be less important. However, for the system geometry used in the numerical calculations in combination with the low energy scales, it may significantly affect the observed behavior. In order to avoid this influence, we chose armchair boundary conditions in the leads that do not provide any edge states. Note in addition, that in realistic systems the zigzag-specific effect would also be suppressed since the zigzag edge state is not protected against disorder when next-nearest neighbor hopping is taken into account~\cite{epub12142}. Therefore, we are convinced that our results based on armchair edges in the reservoirs describe the generic situation for wide leads.

\section{Conclusions and outlook}\label{sec:outlook}

Considering the development of the topic of the Aharonov-Bohm effect in graphene, we can assess that there is a great variety of aspects covered by theoretical considerations, and despite the significant amount of work done on graphene ring systems, there are still a number of open questions drawn from initial experiments that remain unanswered so far, as well as a large number of theoretical predictions not yet confirmed by corresponding experiments. On the one hand, it would be interesting from a theoretical point of view to investigate the origin and significance of the seemingly linear relationship between conductance and oscillation amplitude as well as the significant increase of the oscillation amplitude at high magnetic field observed in Ref.~\onlinecite{PhysRevB.77.085413}. Further analysis on the role of interactions may also be worthwhile. On the other hand, while some of the theoretical models are hardly realizable in experiments at the present stage e.\,g.\ due to insufficient control over edge properties, there are systems, such as presented in the previous two sections, that should be experimentally accessible and robust, and should therefore allow for the observation of graphene-specific features in the Aharonov-Bohm effect.

\begin{acknowledgements}
We acknowledge interesting discussions and collaborations on the subject with C.\,W.\,J.\ Beenakker, Y.\,M.\ Blanter, D.\ Bohr, C.\ Bruder, F.\ Dolcini, K.\ Ensslin, A.\,F.\ Morpurgo, A.\ Rycerz, C.\ Stampfer, and M.\ Wimmer. We further thank the DFG and the ESF for financial support.
\end{acknowledgements}

\appendix



\end{document}